\documentstyle[12pt, bezier]{article}
\def\beq{\begin{equation}}
\def\eeq{\end{equation}}

\def\as{\alpha_s}
\def\ao{{\overline \alpha}}
\def\ed{\varepsilon_d}
\def\eu{\varepsilon_u}
\def\ap#1#2#3 {Ann. Phys. (NY) {\bf#1} (19#2) #3}
\def\apj#1#2#3 {Astrophys. J. {\bf#1} (19#2) #3}
\def\apjl#1#2#3 {Astrophys. J. Lett. {\bf#1} (19#2) #3}
\def\app#1#2#3 {Acta. Phys. Pol. {\bf#1} (19#2) #3}
\def\ar#1#2#3 {Ann. Rev. Nucl. Part. Sci. {\bf#1} (19#2) #3}
\def\cpc#1#2#3 {Computer Phys. Comm. {\bf#1} (19#2) #3}
\def\err#1#2#3 {{\it Erratum} {\bf#1} (19#2) #3}
\def\ib#1#2#3 {{\it ibid.} {\bf#1} (19#2) #3}
\def\jmp#1#2#3 {J. Math. Phys. {\bf#1} (19#2) #3}
\def\ijmp#1#2#3 {Int. J. Mod. Phys. {\bf#1} (19#2) #3}
\def\jetp#1#2#3 {JETP Lett. {\bf#1} (19#2) #3}
\def\jpg#1#2#3 {J. Phys. G. {\bf#1} (19#2) #3}
\def\mpl#1#2#3 {Mod. Phys. Lett. {\bf#1} (19#2) #3}
\def\nat#1#2#3 {Nature (London) {\bf#1} (19#2) #3}
\def\nc#1#2#3 {Nuovo Cim. {\bf#1} (19#2) #3}
\def\nim#1#2#3 {Nucl. Instr. Meth. {\bf#1} (19#2) #3}
\def\np#1#2#3 {Nucl. Phys. {\bf#1} (19#2) #3}
\def\pcps#1#2#3 {Proc. Cam. Phil. Soc. {\bf#1} (#2) #3}
\def\pl#1#2#3 {Phys. Lett. {\bf#1} (19#2) #3}
\def\prep#1#2#3 {Phys. Rep. {\bf#1} (19#2) #3}
\def\prev#1#2#3 {Phys. Rev. {\bf#1} (19#2) #3}
\def\prl#1#2#3 {Phys. Rev. Lett. {\bf#1} (19#2) #3}
\def\prs#1#2#3 {Proc. Roy. Soc. {\bf#1} (19#2) #3}
\def\ptp#1#2#3 {Prog. Th. Phys. {\bf#1} (19#2) #3}
\def\ps#1#2#3 {Physica Scripta {\bf#1} (19#2) #3}
\def\rmp#1#2#3 {Rev. Mod. Phys. {\bf#1} (19#2) #3}
\def\rpp#1#2#3 {Rep. Prog. Phys. {\bf#1} (19#2) #3}
\def\sjnp#1#2#3 {Sov. J. Nucl. Phys. {\bf#1} (19#2) #3}
\def\spj#1#2#3 {Sov. Phys. JEPT {\bf#1} (19#2) #3}
\def\spu#1#2#3 {Sov. Phys.-Usp. {\bf#1} (19#2) #3}
\def\zp#1#2#3 {Zeit. Phys. {\bf#1} (19#2) #3} 

\begin{document}
\begin{titlepage}
\begin{center}
{\Large \bf Theoretical Physics Institute \\
University of Minnesota \\}  \end{center}
\vspace{0.3in}
\begin{flushright}
TPI-MINN-96/1-T \\
UMN-TH-1422-96 \\
February 1996
\end{flushright}
\vspace{0.4in}
\begin{center}
{\Large \bf  Second order QCD corrections to the nonleptonic decay 
of the $b$ quark in the slow charm limit\\}
\vspace{0.2in} 
{\bf M.B. Voloshin  \\ } 
Theoretical Physics Institute, University of Minnesota, Minneapolis, MN 
55455 \\ and \\ 
Institute of Theoretical and Experimental Physics, Moscow, 117259 \\[0.2in]

{\bf   Abstract  \\ }
\end{center}

The perturbative QCD corrections  to the nonleptonic decay rate of the $b$ 
quark are discussed. By considering the limit where the final charmed quarks 
are slow, it is argued that the coefficients of the $\as^2$ terms, 
corresponding to next-to-next-to-leading order in the standard 
renormalization group expansion in $\ln (m_W/m_b)$, are naturally large. The 
large coefficients arise from the final-state gluon exchange between quarks 
and are associated with the region of rather low momenta, which may further 
enhance the phenomenological significance of these terms.  

\end{titlepage}

\section{Introduction}

The problem of the nonleptonic decays of the $B$ mesons is well known and is 
well documented in the literature (see e.g. in \cite{baffling,bbbg}): the 
experimentally measured semileptonic branching ratio $B(B \to l \, \nu \, 
X)$ is too low, and leaves unaccounted in the existing theoretical 
calculations an enhancement of about 20\% or so of the nonleptonic decay 
rate of the $B$ mesons relative to their semileptonic decays. One effect: a 
relative enhancement by possibly as much as 30\% of the sub-dominant decay 
$b \to c \bar c s$ due to the mass of the charm quark in the $O(\as)$ QCD 
correction (which effect is contained in the results of Ref.\cite{hp}, and 
which has been rediscovered more recently$^{\cite{bbbg0,mv,bbfg}}$), is not 
entirely sufficient for the overall enhancement of the total nonleptonic 
rate and also does not look to be supported by the data. Indeed a solution 
of the problem of the semileptonic branching ratio of the $B$ mesons by an 
enhancement of the decay $b \to c \bar c s$ would require an average yield 
of charmed quarks per $B$ decay of about 1.3, whereas the latest 
experimental result from CLEO for this number is$^{\cite{cleo}}$ $n_c=1.15 
\pm 0.044$, which does not support a relative enhancement  of this decay 
mode with respect to the dominant one $b \to c \bar u d$. Thus the 
experimental situation suggests a further theoretical study of the 
nonleptonic $b$ decays beyond the thus far considered terms.

A theoretical study of the ratio of nonleptonic to semileptonic decay rates 
of the $B$ mesons involves two ingredients: perturbative QCD corrections to 
decays of a $b$ quark and the nonperturbative effects related to the fact 
that the $b$ quark decays being confined in a hadron. The latter effects, 
when considered within the expansion in the inverse mass of the $b$ quark, 
are estimated$^{\cite{baffling}}$ to be quite small: of the order of few 
percent. Thus the required by the data enhancement of the nonleptonic decay 
is presumed to be associated with the perturbative QCD effects in the decay 
of the $b$ quark. The standard way of analyzing the perturbative radiative 
corrections in the nonleptonic decays is through the renormalization group 
(RG) summation of the leading log terms and the first next-to-leading 
terms$^{\cite{ap,bbbg}}$ in the parameter $L \equiv \ln(m_W/m_b)$. For the 
semileptonic decays the logarithmic dependence on $m_W/m_b$ is absent in all 
orders due to the weak current conservation at momenta larger than $m_b$, 
thus the correction is calculated by the standard perturbative technique, 
and a complete expression in the first order in $\as$ is available both for 
the total rate$^{\cite{hp,nir}}$ and for the lepton spectrum$^{\cite{cj}}$. 
In reality however the parameter $L \approx 2.8$ is not large, and 
non-logarithmic terms may well compete with the logarithmic ones. This 
behavior is already seen from the known expression for the logarithmic 
terms: when expanded up to the order $\as^2$ the result of Ref.\cite{bbbg0} 
for the rate of decays with single final charmed quark takes the form
\beq
{{\Gamma(b \to c \bar u d) + 
\Gamma(b \to c \bar u s)} \over {3 \, \Gamma(b \to c e
\bar \nu )}}=
 1+ {\as \over \pi} + {\as^2 \over \pi^2} \, \left [ 4 \, L^2 + \left
( {7 \over 6} + { 2 \over 3} \, c(m_c^2/m_b^2) \right ) \, L \right ]~,
\label{as2l}
\eeq
where, in terms of notation of Ref.\cite{bbbg0}, $c(a)=c_{22}(a)-c_{12}(a)$. 
The behavior of the function $c(a)$ is known explicitly$^{\cite{bbbg0}}$ and 
is quite weak:  $c(0)=19/2$, $c(1) = 6$, and $c(m_c^2/m_b^2) \approx 9.0$ 
for the realistic mass ratio $m_c/m_b \approx 0.3$. One can see that the 
term with the single logarithm $L$ contributes about two thirds of that with 
$L^2$ in the term quadratic in $\as$. Under such circumstances the RG 
summation of the terms with powers of $L$ does not look satisfactory for 
numerical estimates of the QCD effects, at least at the so far considered 
level of the first next-to-leading order terms.

An alternative way of evaluating the QCD corrections would be to use the 
straight expansion in the QCD coupling at least to the order $\as^2$, which 
amounts to a calculation of the  non-logarithmic term in the ratio 
(\ref{as2l}) in the order $\as^2$. In terms of the RG summation this 
corresponds to calculating the next-to-next-to-leading terms, which can be 
used for constructing the corresponding full RG expression in that high 
order. Unfortunately an explicit calculation of the $\as^2$ terms poses a 
quite complicated technical problem, which however, perhaps, can eventually 
be solved. The purpose of this paper is to present arguments that the 
non-logarithmic terms in the order $\as^2$ may naturally be large and in 
fact may compete by their numerical significance with the known logarithmic 
terms. In other words the natural parameter for these terms is $\as^2$ in 
the case of $b \to c \bar u d (s)$ decays and even $\pi^2 \as^2$ in the case 
of the decay $b \to c \bar c s$ rather than the $(\as/\pi)^2$ that is usual 
for the radiative effects. To see this the limit of small velocity of the 
$c$ quark in the $b \to c \bar u d (s)$ decays and of small velocity of both 
charmed quarks in the decay $b \to c \bar c s$ is considered (slow charm 
limit).  The realistic kinematics of the $b$ quark decay is 
known$^{\cite{sv}}$ to be not too far from this limit, however it is  far 
enough to prevent from obtaining a reliable quantitative estimates in the 
discussed problem of the real $b$ quark decays\footnote{It can be also 
noticed that for the decay $b \to c \bar c s$ the inverse of the velocity 
$v$ of either of the charmed quarks in the center of mass of the $c \bar c$ 
system, averaged over the phase space in the decay, is $\langle v^{-1} 
\rangle \approx 2$. Thus $v^{-1}$ is not a much worse parameter than $L$. 
However, as will be seen in the real kinematics, an expansion in $v$ of the 
$\as^2$ terms lacks convergence similarly to the expansion in $L$.}. The 
reason for the large coefficients, arising in the limit of slow charm is the 
Coulomb-like gluonic interaction of the quarks in the final state. The 
presence of this interaction makes the nonleptonic $b$ quark decays 
fundamentally different from the semileptonic ones at least in the slow 
charm limit.  The QCD corrections to the semileptonic decay are finite in 
this limit and are determined by exchange of virtual gluons with the scale 
of momenta between $m_c$ and $m_b\, ^{\cite{sv}}$. On the contrary, the QCD 
corrections to the nonleptonic decay rate are singular in the limit of zero 
velocity of the final charmed quark already in the coefficients of the 
$\as^2$ terms.

The set of essential graphs, which need to be calculated for a complete 
$O(\as^2)$ QCD analysis of the corrections to nonleptonic decays of $b$, is 
somewhat simplified$^{\cite{mv}}$ for the ratio $\Gamma_{nl}/\Gamma_{sl}$, 
as will be discussed in Sect.2. This simplification allows to ignore the 
non-Abelian nature of the QCD interaction and to relate the problem to the 
known solution of a Coulomb problem for non-relativistic as well as for 
relativistic fermions. In Sect.3 the relativistic case is discussed, 
relevant for the decay $b \to c \bar u d(s)$ and in Sect.4 the decay $b \to 
c \bar c s$ is considered, where both the relativistic and non-relativistic 
Coulomb dynamics comes into play.  In Sect.5 the results of this paper are 
summarized and discussed.

\section{Feynman graphs for $\Gamma_{nl}/\Gamma_{sl}$}

The structure of the graphs, which need to be calculated in order to find 
the QCD corrections to $\Gamma_{nl}/\Gamma_{sl}$ up to $O(\as^2)$, becomes 
transparent if one represents the decay rate as the imaginary part of the 
$b$ quark self-energy. Then the zeroth order nonleptonic decay rate is given 
by the graph of Fig.1, while the $O(\as)$ and the $O(\as^2)$ corrections are 
represented by the classes of graphs shown respectively in Figures 2 and 
3\footnote{We neglect here the penguin type contributions, originating from 
the process $b \to c \bar c s$, since these are known and 
small$^{\cite{ap}}$.}.  It is convenient for the purpose of discussion to 
call the quark line $b \to c \to b$ in those graphs as the heavy quark line, 
and the loop, made of $\bar u d(s)$ or $\bar c s$ quarks as the light quark 
loop. The graphs, where the gluons are attached only to the heavy quark line 
(like those shown in Fig.2a and Fig.3a), are identical to the same graphs 
for the semileptonic decays and thus they cancel in the ratio 
$\Gamma_{nl}/\Gamma_{sl}$. Therefore the corrections to this ratio should 
contain either gluon exchange within the light quark loop, or a gluon 
exchange between the loop and the heavy quark line, or both.  Notice however 
that a color trace is taken over the light quark loop, thus the graphs where 
an overall color goes into the loop do not contribute to the decay rate 
(examples of such graphs are in Fig.2c and also in Fig.4). Therefore in the 
first order in $\as$ only the graph of Fig.2b is left with the gluon 
exchange within the loop. For light quarks this exchange is very well known 
to result in the factor $(1+\as/\pi)$, which enters eq.(\ref{as2l}), while 
for the $\bar c s$ loop the charm quark mass enhances the correction, so 
that for realistic quark masses the correction factor 
is$^{\cite{bbbg0,mv,bbfg}}$ $(1+\delta \, \as/\pi)$ with $\delta \approx 
4.5$.

In the order $\as^2$ the graphs for the corrections to the ratio 
$\Gamma_{nl}/\Gamma_{sl}$ fall into two categories: QCD corrections in the 
order $\as^2$ entirely contained within the loop and a two-gluon exchange 
between the heavy quark line and the loop. Due to the color trace over the 
loop there is no interference in this order of the gluon exchange between 
the heavy quark line and the loop and either the corrections inside the loop 
or on the heavy quark line. Also the gluon self-energy, or non-Abelian 
splitting of the gluon into two for the gluon exchanged between the heavy 
line and the loop does not contribute due to the color trace. Thus the only 
exchange graphs that are relevant are those with two gluons, where both 
gluons originate on the heavy quark line and end on the quarks within the 
loop. An example of such graph is shown in Fig.3c. 

The second order QCD corrections contained within the loop are well
known for the massless quarks$^{\cite{ckt,ds}}$. 
They replace the first-order factor $(1+\as/\pi)$ by
\begin{eqnarray}
1+{\as^{\overline {MS}}(q) \over \pi} + 
\left [ \left ( {41 \over 8} -{11 \over 3} \zeta(3) \right ) \, C_A 
\right. & - & \left. {1
\over 8} \, C_F + \left ( -{11 \over 12} + {2 \over 3} \zeta(3) \right )
\, n_f \right] \, \left ( {\as^{\overline {MS}}(q) \over \pi} \right )^2
\approx 
\nonumber \\
1+{\as^{\overline {MS}}(q) \over \pi} & + &
1.64 \left ( {\as^{\overline {MS}}(q) \over \pi}\right )^2
\label{loop2}
\end{eqnarray}
with the number of light quark flavors $n_f=3$ and $q^2$ being the 
square of the invariant mass of the quarks in the loop. Thus in this
case the effect of the second order basically reduces to specifying the
normalization scale for the coupling $\as$ in the first-order term and
does not contain any spectacular features beyond that. For the $\bar c
s$ loop, where the mass of the charmed quark cannot be neglected, the
corresponding result for the QCD correction in order $\as^2$ is not
known.

Proceeding to the discussion of the graphs with two-gluon exchange between 
the heavy quark line and the light quark loop, we first notice that such 
graphs have no non-Abelian structure and their contribution is exactly the 
same as it would be in an Abelian theory. The only feature of the color 
structure of QCD that is relevant is the color factor originating from the 
color traces. One can readily see that to correctly reproduce the effect of 
the two-gluon exchange between the heavy line and the loop relative to the 
zeroth order contribution of the diagram of Fig.1, one can replace the QCD 
interaction by an Abelian theory with the coupling ${\overline \alpha}$ 
related to $\as$ as ${\overline \alpha}^2 = 2 \as/9$. (This takes into 
account the color factor of 3 present in the graph of Fig.1). 

A part of the contribution of these graphs is contained in the $\as^2$ term 
in eq.(\ref{as2l}), that is the part, which is due to the exchange of 
virtual gluons with momenta between $m_b$ and $m_W$. Here we will be 
concerned with the opposite part of the spectrum, i.e. with gluons with 
momenta much less than $m_c$. In general, the imaginary part of graphs of 
the type shown in Fig.3c is contributed by exchange of virtual gluons and by 
radiation and absorption of real transversal gluons. However, for a slow 
charmed quark its radiation and absorption arises only in the order $v^2$ of 
the expansion in its velocity\footnote{Notice, that the graphs where a gluon 
is emitted and absorbed by a light quark are not of the type of Fig.3c. In 
the graphs of this type at least one emission or absorption has to be by a 
heavy quark.}.  Thus, the leading terms are determined only by the exchange 
of virtual gluons. Furthermore, it is only the spin-independent (for heavy 
quark) exchange of Coulomb gluons that contributes in the leading order for 
a slow quark. Therefore the problem reduces to the one-particle problem of 
motion of a quark in an Abelian Coulomb-like field of a heavy quark. For the 
non-relativistic motion of the two charmed quarks in the decay $b \to c \bar 
c s$ this is the standard Quantum Mechanical problem, while for the light 
quarks both in this decay and in $b \to c \bar u d(s)$ the problem reduces 
to the known solution of the Dirac equation in a Coulomb field (see e.g. in 
the textbook \cite{blp}). In the former case the singularity in the velocity 
of the charmed quarks has the well known form $\pi^2 \as^2/v^2$, while in 
the latter case a logarithmic singularity arises due to the corresponding 
singular behavior of the relativistic Coulomb wave function at the origin in 
the order $\as^2$.

\section{The decay $b \to c \bar u d(s)$.}

We first consider the decay $b \to c \bar u d$, where both quarks recoiling 
against the charmed quark are genuinely light (the decay $b \to c \bar u s$ 
is obviously similar in as much as the masses of both the $d$ and $s$ quarks 
can be neglected). As described above, this decay is considered in the limit 
of slow charm, i.e. in the limit $\Delta \equiv m_b -m_c \ll m_b,m_c$. In 
this limit the final charmed quark is considered to be at rest, and, as it 
will be shown, the $\as^2$ term in the correction to the decay rate develops 
a coefficient proportional to $\ln(m_c/\Delta)$. Certainly, the decay rate 
itself goes to zero in the limit $\Delta \to 0$ due to the overall 
$\Delta^5$ factor, and it is only the ratio of the $\as^2$ correction term 
to the zeroth order decay rate that displays a logarithmic singularity in 
this limit. Also as explained above, in order to find the soft gluon 
contribution in the two-gluon exchange graphs the QCD interaction can be 
replaced by an Abelian one with the coupling $\ao^2=2 \as^2/9$. The 
interaction of the final light quarks with the charmed quark, acting as a 
Coulomb center, is taken into account by replacing the plane wave functions 
for the light quarks by their wave functions in the Coulomb field. Thus we 
briefly discuss an adaptation of the known textbook results$^{\cite{blp}}$ 
to the case considered here.  Before proceeding to this discussion, let us 
note that in the calculation of the Coulomb wave functions of the light 
quarks the charmed quark can be considered as a static center located at 
${\bf r}=0$, only for the wave functions at the distances $ r \gg r_0 
\approx m_c^{-1}$, since at shorter distances the momenta of the exchanged 
gluons are not small in comparison with the mass of the charmed quark, and 
its recoil cannot be neglected. Thus we will use the light quark wave 
functions with the logarithmic accuracy down to these distances in the weak 
interaction Hamiltonian proportional to $(u^\dagger(r_0) \, (1-\gamma_5) \, 
d(r_0))$.

The Dirac-Weyl equation for a massless two-component left-handed $d$ quark 
spinor in the repulsive Coulomb field $U(r)=\ao/r$ reads as
\beq
\left (\ed -U(r) - i \, \mbox{\boldmath $\sigma \cdot  \nabla$} \right) \,
d(\bf r) =0~~,
\label{dw}
\eeq
where $\ed$ is the energy of the $d$ quark. A non-vanishing near the
origin solution to this equation arises only in the lowest partial wave,
where the ansatz for the spinor structure of the solution has the form
\beq
d({\bf r})=\left ( f(r) +i \, g(r) \, \mbox{\boldmath $\sigma \cdot n$}
\right)\, w
\label{anz}
\eeq
with $w$ being a constant two-component spinor, and ${\bf n}= {\bf
r}/r$. Upon substitution of this ansatz in eq.(\ref{dw}) the resulting
equations for the functions $f(r)$ and $g(r)$ exactly coincide with the
textbook ones$^{\cite{blp}}$ in the massless case $m=0$ and the angular
moment parameter $\kappa=-1$. Thus the solution can be readily read off
the textbook:
\beq
\left. \begin{array}{c}
           f \\ g
         \end{array} 
\right \} = 2^{3 \over 2} e^{-{\pi \ao \over 2}} { \Gamma
(\gamma + 1- i \, \ao) \over \Gamma(2 \gamma +1)} { (2 \, p \,
r)^\gamma \over r} \, 
\begin{array}{c}
          {\rm Im} \\ {\rm Re}
\end{array} \left \{ e^{i(pr+\xi)}\, _1F_1(\gamma + i \, \ao, 2\gamma +1,
-2ipr) \right \}~~,
\label{sol}
\eeq
where $\gamma = \sqrt{1-\ao^2}$, $p=\ed$, and the phase angle $\xi$ is 
defined as $\exp(-2i\, \xi)=-\gamma-i\ao$.  The wave function of the right 
handed $\bar u$ antiquark is obtained from this solution by a formal 
replacement $\ao \to -\ao$, $g \to -g$. 

In calculating the matrix element of the weak interaction Hamiltonian we are 
interested in these functions at a distance $r_0$ such that $p r_0 \ll 1$, 
but still $p r_0 \gg \ao$, since we are seeking an expansion in powers of 
$\ao$ and if the latter condition were not satisfied such an expansion would 
be impossible. Due to the former condition the degenerate hypergeometric 
function can be replaced by 1. Multiplying the solutions to the wave 
equations in the expression $(u^\dagger (r_0) d(r_0))$, one finds the ratio 
of the Coulomb-corrected matrix element to the bare one in the following 
form
\beq
{\left(u^\dagger (r_0)d(r_0)\right)_{Coulomb} \over \left(u^\dagger
(0)d(0)\right)_{free}} = \left | { 2\, \Gamma(\gamma + 1 + i \, \ao)
 \over \Gamma(2 \gamma +1)} \right |^2 (4 \,\eu \, \ed r_0^2)^{\gamma-1}~.
\label{rat}
\eeq
One can readily see that the parameter for the expansion of this
expression is $\ao^2$ rather than $(\ao/\pi)^2$. Under the assumptions
made we are however bound to retain only the logarithmic part of this
expression and only in the order $\ao^2$. With the logarithmic accuracy
one can set $r_0 \sim m_c^{-1}$, $\eu \sim \ed \sim \Delta$ thus finding
for the correction to the rate of the decay, given by the square of the
ratio in eq.(\ref{rat}), in the form of the factor
\beq
F_{\bar u d}= \left ( 1+ {4 \over 9} \, \as^2 \, \ln {m_c \over \Delta}
\right )~~,
\label{udres}
\eeq
where the relation $\ao^2 = 2 \as^2/9$ is used. This formula is the
final result for the logarithmic singularity of the $\as^2$ terms in the
decay $b \to \bar u d(s)$ in the slow charm limit of $\Delta \to 0$.

Clearly, for the realistic $b$ quark decay
the equation (\ref{udres}) can serve only as an indicator of
possible presence of large terms of order $\as^2$, rather than as a
quantitative estimate of these terms, since in reality the charm in $b$
decay is not sufficiently slow for considering $\ln(m_c/\Delta)$ as a
large parameter. It is interesting to note however, that the equation
(\ref{rat}) can be used to find the exact expression for the $\as^2$
correction to a heavy quark decay in an entirely artificial kinematical
situation, where both the $b$ and $c$ quarks are heavier than the $W$
boson: $\Delta \ll m_W \ll m_b,\, m_c$\,\footnote{Actually the condition
$\Delta \ll m_W$ can be dropped. It is assumed here in order to simplify
the resulting expression, which anyway serves only for the purpose of
illustration.}. In this limit the $W$ boson propagator provides a cutoff
at short distances $r_0 \sim m_W^{-1}$, at which the recoil of the heavy
quark can still be ignored, thus the one-particle wave functions of the
form in eq.(\ref{sol}) can be used for a complete calculation of the
matrix element of the weak Hamiltonian. This matrix element is
proportional to 
\beq
J=\int (u^\dagger ({\bf r}) d({\bf r}))\, 
{\exp(-m_W \, r) \over r} \, d^3r
\label{j}~.
\eeq 
Calculating this integral with the solutions described by
eq.(\ref{sol}), one readily finds for the square of the 
ratio of the Coulomb-corrected
integral to the free one the following expression
\beq
\left | {J(\eu, \, \ed)_{Coulomb} \over J(\eu, \, \ed)_{free}} 
\right | ^2 =
\Gamma(2\gamma)^2 \, \left | { 2\, \Gamma(\gamma + 1 + i \, \ao)
 \over \Gamma(2 \gamma +1)} \right |^4 \left ( {4 \, \eu \, \ed \over
m_W^2 } \right )^{2 \gamma -2} \, \left [ 1 + O 
\left ( {\Delta^2 \over m_W^2} \right) \right ] ~~.
\label{ratj}
\eeq
Since the phase space integral with the free wave functions goes in the
considered kinematical arrangement as $\eu^2 \, \ed^2 \, d\ed$ with
$\eu=\Delta - \ed$, one finds the Coulomb correction factor
for the decay rate as
\begin{eqnarray}
{\cal F} & = & \Gamma(2\gamma)^2 \, { 30 \, \Gamma(2 \gamma +1)^2 \over
\Gamma(4\gamma+2)} \,
\left | { 2\, \Gamma(\gamma + 1 + i \, \ao)
 \over \Gamma(2 \gamma +1)} \right |^4 \left ( 4 \Delta^2 \over m_W^2
\right )^{2\gamma -2} \nonumber \\
&=& {120 \over \gamma^2 \, \Gamma(4 \gamma +2)} \, \left | \Gamma
(\gamma + i \, \ao) \right |^4 \,\left ( 4 \Delta^2 \over m_W^2
\right )^{2\gamma -2}
\label{fcoul}
\end{eqnarray}
(the identity $\gamma^2+\ao^2=1$ is used in the last transition).
When expanded to the order $\ao^2$, this factor takes the form
\beq
F = 1 + \left ( {167 \over 30} - {\pi^2 \over 3} + 2 \ln {m_W \over 2 \,
\Delta} \, \right ) \, {2 \over 9} \, \as^2~.
\label{artf}
\eeq
This expression describes both the logarithmic term in the $\as^2$
correction and the non-logarithmic one, which is justified to be
retained in the assumed artificial limit $\Delta \ll m_W \ll m_b,\,
m_c$. When combined with the result in eq.(\ref{loop2}) for the QCD 
corrections contained within the light quark loop, this would
completely describe the QCD corrections up to the order $\as^2$ in this
limit. It can be also mentioned that in this limit there of course are
no terms with $\ln(m_W/m_b)$.

\section{The decay $b \to c \bar c s$}

In the decay $b \to c \bar c s$ in the slow charm limit one
has two slow quarks in the final state. An exchange of gluons between
them results in the well-known non-relativistic Coulomb corrections,
which have as their parameter $\pi \as/v$, where $v$ is the velocity of
either of the charmed quarks in their center of mass frame:
$v=\sqrt{1-4m_c^2/q^2}$ with $q^2$ being the invariant mass squared of the 
$c \bar c$ pair. Thus in the limit of small $v$ the dominant $\as^2$ term
is given by the double Coulomb exchange between the $c$ and $\bar c$,
and is described by the second term of the expansion of the Coulomb
factor 
\beq
{\pi \ao /v \over 1-\exp(-\pi \ao /v)} = 1+ {\pi \ao \over v} + { \pi^2
\ao^2 \over 12 \, v^2} + \ldots
\label{cf}
\eeq
The linear term in $\ao$ should be discarded, as explained in Sect.2, 
while the second term gives the correction factor 
\beq
1+ \pi^2 \as^2 /(54
v^2)
\label{v2cf}
\eeq
for the rate (the relation $\ao^2=2\as^2/9$ is again used here, 
which takes into account the color factors in the relative correction).

However, one in fact can estimate at least one more term in the
expansion of the $\as^2$ correction in powers of $v$, and thus get some
feeling of what $v$ is sufficiently small for applicability of the
expansion. This is especially relevant, given rather small numerical
coefficient in the leading correction term in eq.(\ref{v2cf}).
The next term, linear in $1/v$ arises through the
interference of the one Coulomb gluon exchange between $c$ and $\bar c$,
described by the term $O(\ao)$ in eq.(\ref{cf}) with the less singular
than $1/v$ part of the gluon exchange of either the $\bar c$ or $s$
quarks with the heavy quark line. For the $\bar c$ this corresponds to
a hard gluon exchange with either the $c$ or the $b$ quark,
while for the $s$ quark  this corresponds to contribution of gluons with
generally arbitrary momenta. The contribution of the soft gluons for the
$s$ quark exchange with the heavy quark line is however not singular at
all in $v$: the logarithmic singularity develops only starting from the
order $\ao^2$, as described by the wave function in eq.(\ref{sol}).
Therefore the linear in $1/v$ term 
has no additional logarithmic dependence
on $v$. The non-logarithmic coefficient can be estimated in the same
approximation as in eq.(\ref{as2l}) i.e. using $L$ as a parameter. In
this approximation the exchange of the non-Coulomb gluon is dominated by
the region of momenta between $m_b$ and $m_W$, and one finds that the
interference of the hard gluon exchange with a Coulomb exchange within
the $c \bar c$ pair adds a negative contribution to the correction
factor (\ref{v2cf}): $-3 \, {\ao^2 \over v} \, L = - {2 \over 3} \,
{\as^2 \over v} \, L$. Thus the final result for the $\as^2$ correction
factor to the $b \to c \bar c s$ decay rate, unaccounted for by
eq.(\ref{as2l}), in the slow charm limit is estimated as
\beq
F_{\bar c s}= 1+ \left ( {\pi^2 \over 54 \, v^2} - {2 \over 3 v}\,
(L+O(1))\right ) \, \as^2~~.
\label{fcs}
\eeq

The formula in eq.(\ref{fcs}) describes the leading $\as^2$ terms in the
part of the spectrum, where the $c \bar c$ pair is close to its 
threshold. In
the total decay rate of the real $b$ quark the dominance of the singular 
in $v$ terms is however quite smeared. Indeed, for the average over the
spectrum of the $b \to c \bar c s$ decay values of $v^{-2}$ and $v^{-1}$
one finds at $m_c/m_b \approx 0.3$:
\beq
\langle v^{-2} \rangle \approx 5.14~,~~~~~~~~
\langle v^{-1} \rangle \approx 1.97~~.
\label{avv}
\eeq
Thus the linear in $1/v$ term in eq.(\ref{fcs}) in fact dominates over
the quadratic one and makes the overall coefficient of $\as^2$ negative
and large: $F_{\bar c s} \approx 1-2.8 \, \as^2$. Certainly, under these
circumstances one can not rely on the expansion in $v$ for the total
rate and may only consider eq.(\ref{fcs}) as an indication of presence
of large contributions in the order $\as^2$. 

If the coefficients of the $\as^2$ terms may be as big as indicated by
the present analysis, these terms may well compete with the first-order
term in $\as$. Thus the $\as^2$ effects may essentially modify the
conclusion$^{\cite{bbbg0,mv,bbfg}}$ based on the first order in $\as$
calculation about the relative enhancement of the $b \to c \bar c s$
decay.

It should be also noted in connection with the $b \to c \bar c s$ decay
that the gluon exchange contained within the $\bar c s$ loop receives a
logarithmic contribution near the threshold in the order $\as^2$ from
the Coulomb exchange. This contribution also can be readily found from
eq.(\ref{sol}). However in this case one has a competing logarithmic
effect due to the so-called hybrid$^{\cite{sv2}}$ anomalous dimension of
the current $(\bar s_L \, \gamma_\mu \, c)$ both in the leading order
and in the next-to-leading order\footnote{The known enhancement of 
this decay in the first order in $\as$  is partly due to the logarithmic
rise of the $\as$ term towards the $\bar c s$ threshold$^{\cite{mv}}$.}.
In any event, however, these terms are less singular in the slow charm
limit than those in eq.(\ref{fcs}).

\section{Summary and discussion}

Lacking a complete calculation of the second-order QCD corrections to
the $b$ quark decays, we have to rely on parametric approximations for
these corrections. Unfortunately, in the real situation there is no such
parameter, which would be acceptably good. The widely used expansion of 
these corrections in $L=\ln(m_W/m_b) \approx 2.8$ by means of RG
does not seem to work
well, as illustrated by eq.(\ref{as2l}). In this paper a different
parameter is considered: the velocity of the charmed quarks in the final
state of the decay. The terms which are parametrically large in the limit
of large $1/v$ are in general independent of those, which dominate the
expansion in $L$, thus they indicate the magnitude of the contributions
lost in the so-far considered order of the expansion in $L$. Also in the
decay $b \to c \bar c s$ the inverse velocity of the charmed quarks is
almost as good (or as bad) a parameter as $L$. 
As is shown the leading in the slow charm limit the $\as^2$ corrections 
are associated with the interaction of
the final state quarks with the Coulomb-like gluonic field of the slow
quark. Therefore the parameter for these corrections turns out to be
$\as^2$ in the case of the decay $b \to c \bar u d (s)$ with
relativistic light quarks (eq.(\ref{udres})) or $(\pi \as)^2$ for the
case of the interaction between the non-relativistic 
$c$ and $\bar c$ in the
decay $b \to c \bar c s$ (eq.(\ref{fcs})). This can be considered as a
strong indication that the missing in eq.(\ref{as2l}) terms of order
$\as^2$ can have large coefficients, which may be considerably different
in the $b \to c \bar u d (s)$ decay from those in the $b \to c \bar c s$
decay. In this case the missing contributions may well compete in their
numerical significance with the terms present in eq.(\ref{as2l}) and may
considerably alter the theoretical predictions both for the 
semileptonic branching ratio  and for the average charm
yield in $B$ decays. Thus the main conclusion from the arguments
presented in this paper is that it is rather premature to seek a
contradiction of the theoretical predictions with the data on $B_{sl}$
and $n_c$ before a complete theoretical analysis of the $\as^2$
corrections becomes available.

This work is supported, in part, by the DOE grant DE-AC02-83ER40105.

\newpage

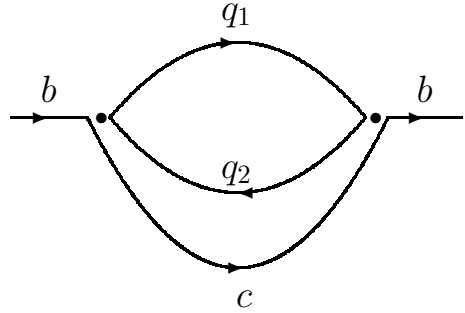
\begin{figure}
\thicklines
\unitlength=1.00mm
\begin{picture}(104.00,51.00)
\put(44.00,31.00){\vector(1,0){5.00}}
\put(49.00,31.00){\line(1,0){5.00}}
\put(94.00,31.00){\vector(1,0){5.00}}
\put(99.00,31.00){\line(1,0){5.00}}
\bezier{356}(54.00,31.00)(74.00,-9.00)(94.00,31.00)
\bezier{208}(57.00,31.00)(74.00,11.00)(91.00,31.00)
\bezier{208}(57.00,31.00)(74.00,51.00)(91.00,31.00)
\put(73.00,41.00){\vector(1,0){1.00}}
\put(75.00,21.00){\vector(-1,0){1.00}}
\put(73.00,11.00){\vector(1,0){2.00}}
\put(49.00,33.00){\makebox(0,0)[cb]{\large $b$}}
\put(99.00,33.00){\makebox(0,0)[cb]{\large $b$}}
\put(75.00,8.00){\makebox(0,0)[ct]{\large $c$}}
\put(74.00,44.00){\makebox(0,0)[cb]{\large $q_1$}}
\put(74.00,23.00){\makebox(0,0)[cb]{\large $q_2$}}
\put(56.00,31.00){\circle*{1.50}}
\put(92.50,31.00){\circle*{1.50}}
\end{picture}
\caption{The lowest order graph, whose unitary cut describes the rate
of the decay $b \to c \bar q_2 q_1$. The small filled circles represent
the W boson propagators.}
\end{figure}

\thicklines
\unitlength=1.00mm
\begin{figure}
\begin{picture}(140.00,100.00)
\put(10.00,80.00){\vector(1,0){5.00}}
\put(15.00,80.00){\line(1,0){5.00}}
\put(60.00,80.00){\vector(1,0){5.00}}
\put(65.00,80.00){\line(1,0){5.00}}
\bezier{356}(20.00,80.00)(40.00,40.00)(60.00,80.00)
\bezier{208}(23.00,80.00)(40.00,60.00)(57.00,80.00)
\bezier{208}(23.00,80.00)(40.00,100.00)(57.00,80.00)
\put(39.00,90.00){\vector(1,0){1.00}}
\put(41.00,70.00){\vector(-1,0){1.00}}
\put(39.00,60.00){\vector(1,0){2.00}}
\put(15.00,82.00){\makebox(0,0)[cb]{\large $b$}}
\put(65.00,82.00){\makebox(0,0)[cb]{\large $b$}}
\put(41.00,57.00){\makebox(0,0)[ct]{\large $c$}}
\put(40.00,93.00){\makebox(0,0)[cb]{\large $q_1$}}
\put(40.00,72.00){\makebox(0,0)[cb]{\large $q_2$}}
\put(22.00,80.00){\circle*{1.50}}
\put(58.50,80.00){\circle*{1.50}}
\put(80.00,80.00){\vector(1,0){5.00}}
\put(85.00,80.00){\line(1,0){5.00}}
\put(130.00,80.00){\vector(1,0){5.00}}
\put(135.00,80.00){\line(1,0){5.00}}
\bezier{356}(90.00,80.00)(110.00,40.00)(130.00,80.00)
\bezier{208}(93.00,80.00)(110.00,60.00)(127.00,80.00)
\bezier{208}(93.00,80.00)(110.00,100.00)(127.00,80.00)
\put(109.00,90.00){\vector(1,0){1.00}}
\put(111.00,70.00){\vector(-1,0){1.00}}
\put(109.00,60.00){\vector(1,0){2.00}}
\put(85.00,82.00){\makebox(0,0)[cb]{\large $b$}}
\put(135.00,82.00){\makebox(0,0)[cb]{\large $b$}}
\put(111.00,57.00){\makebox(0,0)[ct]{\large $c$}}
\put(110.00,93.00){\makebox(0,0)[cb]{\large $q_1$}}
\put(110.00,72.00){\makebox(0,0)[cb]{\large $q_2$}}
\put(92.00,80.00){\circle*{1.50}}
\put(128.50,80.00){\circle*{1.50}}
\put(45.00,35.00){\vector(1,0){5.00}}
\put(50.00,35.00){\line(1,0){5.00}}
\put(95.00,35.00){\vector(1,0){5.00}}
\put(100.00,35.00){\line(1,0){5.00}}
\bezier{356}(55.00,35.00)(75.00,-5.00)(95.00,35.00)
\bezier{208}(58.00,35.00)(75.00,15.00)(92.00,35.00)
\bezier{208}(58.00,35.00)(75.00,55.00)(92.00,35.00)
\put(74.00,45.00){\vector(1,0){1.00}}
\put(76.00,25.00){\vector(-1,0){1.00}}
\put(74.00,15.00){\vector(1,0){2.00}}
\put(50.00,37.00){\makebox(0,0)[cb]{\large $b$}}
\put(100.00,37.00){\makebox(0,0)[cb]{\large $b$}}
\put(76.00,12.00){\makebox(0,0)[ct]{\large $c$}}
\put(75.00,48.00){\makebox(0,0)[cb]{\large $q_1$}}
\put(75.00,27.00){\makebox(0,0)[cb]{\large $q_2$}}
\put(57.00,35.00){\circle*{1.50}}
\put(93.50,35.00){\circle*{1.50}}
\multiput(102.00,72.00)(0,3.50){5}{\line(0,1){2.00}}
\multiput(29.00,66.00)(3.35,0){7}{\line(1,0){1.90}}
\multiput(70.00,16.00)(0,2.80){4}{\line(0,1){1.60}}
\put(40.00,52.00){\makebox(0,0)[ct]{{\Large a}}}
\put(112.00,53.00){\makebox(0,0)[ct]{{\Large b}}}
\put(76.00,6.00){\makebox(0,0)[ct]{{\Large c}}}
\end{picture}
\caption{Three types of graphs, whose unitary cuts describe the first QCD 
radiative corrections to the inclusive decay rate $b \to c \overline q_2 
q_1$. The 
dashed lines correspond to gluons. The gluon vertices can be anywhere on the 
$bc$ line (a), quark lines in the loop (b), or one vertex anywhere on the 
$bc$ line and the other vertex on either line in the loop (c). The
graphs of the type $c$ in fact do not contribute to the decay rate, as
explained in the text.} 
\end{figure}
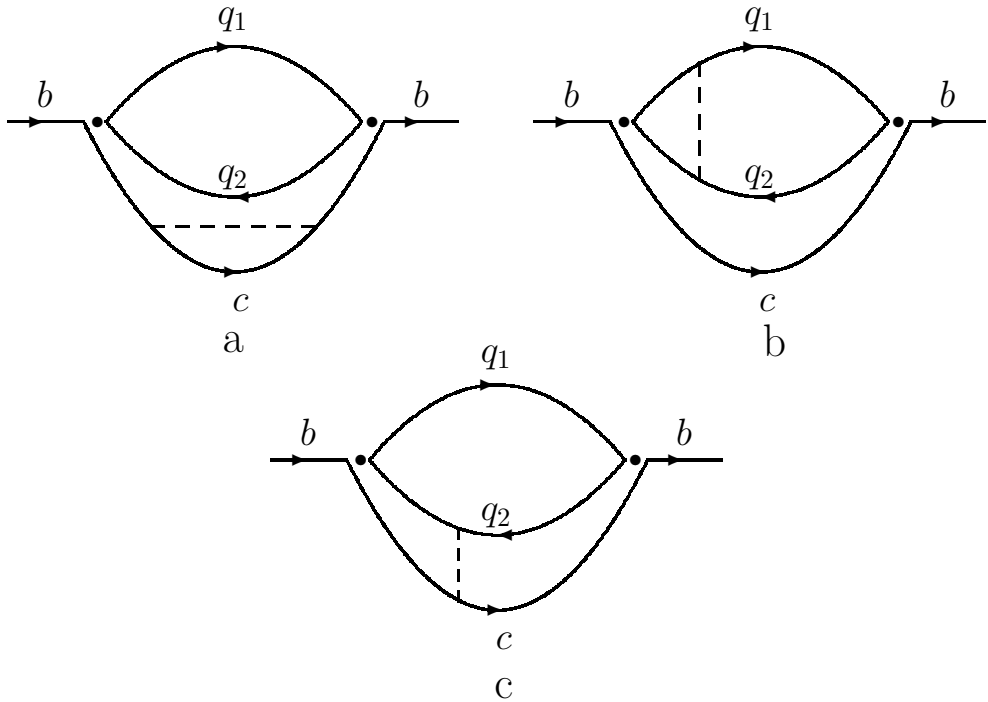

\begin{figure}
\begin{picture}(140.00,100.00)
\put(10.00,80.00){\vector(1,0){5.00}}
\put(15.00,80.00){\line(1,0){5.00}}
\put(60.00,80.00){\vector(1,0){5.00}}
\put(65.00,80.00){\line(1,0){5.00}}
\bezier{356}(20.00,80.00)(40.00,40.00)(60.00,80.00)
\bezier{208}(23.00,80.00)(40.00,60.00)(57.00,80.00)
\bezier{208}(23.00,80.00)(40.00,100.00)(57.00,80.00)
\put(39.00,90.00){\vector(1,0){1.00}}
\put(41.00,70.00){\vector(-1,0){1.00}}
\put(39.00,60.00){\vector(1,0){2.00}}
\put(15.00,82.00){\makebox(0,0)[cb]{\large $b$}}
\put(65.00,82.00){\makebox(0,0)[cb]{\large $b$}}
\put(41.00,57.00){\makebox(0,0)[ct]{\large $c$}}
\put(40.00,93.00){\makebox(0,0)[cb]{\large $q_1$}}
\put(40.00,72.00){\makebox(0,0)[cb]{\large $q_2$}}
\put(22.00,80.00){\circle*{1.50}}
\put(58.50,80.00){\circle*{1.50}}
\put(80.00,80.00){\vector(1,0){5.00}}
\put(85.00,80.00){\line(1,0){5.00}}
\put(130.00,80.00){\vector(1,0){5.00}}
\put(135.00,80.00){\line(1,0){5.00}}
\bezier{356}(90.00,80.00)(110.00,40.00)(130.00,80.00)
\bezier{208}(93.00,80.00)(110.00,60.00)(127.00,80.00)
\bezier{208}(93.00,80.00)(110.00,100.00)(127.00,80.00)
\put(109.00,90.00){\vector(1,0){1.00}}
\put(111.00,70.00){\vector(-1,0){1.00}}
\put(109.00,60.00){\vector(1,0){2.00}}
\put(85.00,82.00){\makebox(0,0)[cb]{\large $b$}}
\put(135.00,82.00){\makebox(0,0)[cb]{\large $b$}}
\put(111.00,57.00){\makebox(0,0)[ct]{\large $c$}}
\put(110.00,93.00){\makebox(0,0)[cb]{\large $q_1$}}
\put(110.00,72.00){\makebox(0,0)[cb]{\large $q_2$}}
\put(92.00,80.00){\circle*{1.50}}
\put(128.50,80.00){\circle*{1.50}}
\put(45.00,35.00){\vector(1,0){5.00}}
\put(50.00,35.00){\line(1,0){5.00}}
\put(95.00,35.00){\vector(1,0){5.00}}
\put(100.00,35.00){\line(1,0){5.00}}
\bezier{356}(55.00,35.00)(75.00,-5.00)(95.00,35.00)
\bezier{208}(58.00,35.00)(75.00,15.00)(92.00,35.00)
\bezier{208}(58.00,35.00)(75.00,55.00)(92.00,35.00)
\put(74.00,45.00){\vector(1,0){1.00}}
\put(76.00,25.00){\vector(-1,0){1.00}}
\put(74.00,15.00){\vector(1,0){2.00}}
\put(50.00,37.00){\makebox(0,0)[cb]{\large $b$}}
\put(100.00,37.00){\makebox(0,0)[cb]{\large $b$}}
\put(76.00,12.00){\makebox(0,0)[ct]{\large $c$}}
\put(75.00,48.00){\makebox(0,0)[cb]{\large $q_1$}}
\put(75.00,27.00){\makebox(0,0)[cb]{\large $q_2$}}
\put(57.00,35.00){\circle*{1.50}}
\put(93.50,35.00){\circle*{1.50}}
\multiput(102.50,72.00)(0,3.50){5}{\line(0,1){2.00}}
\multiput(29.00,66.00)(3.35,0){7}{\line(1,0){1.90}}
\multiput(70.00,16.00)(0,2.80){4}{\line(0,1){1.60}}
\put(40.00,52.00){\makebox(0,0)[ct]{{\Large a}}}
\put(112.00,53.00){\makebox(0,0)[ct]{{\Large b}}}
\put(76.00,6.00){\makebox(0,0)[ct]{{\Large c}}}
\multiput(33.8,62.10)(3.5,0){4}{\line(1,0){2.10}}
\multiput(118.00,72.00)(0,3.50){5}{\line(0,1){2.00}}
\multiput(80.50,16.00)(0,2.80){4}{\line(0,1){1.60}}
\end{picture}
\caption{Three types of graphs, whose unitary cuts contribute to $O(\as^2)$ 
QCD radiative corrections to the inclusive decay rate $b \to c \overline q_2 
q_1$. } 
\end{figure}
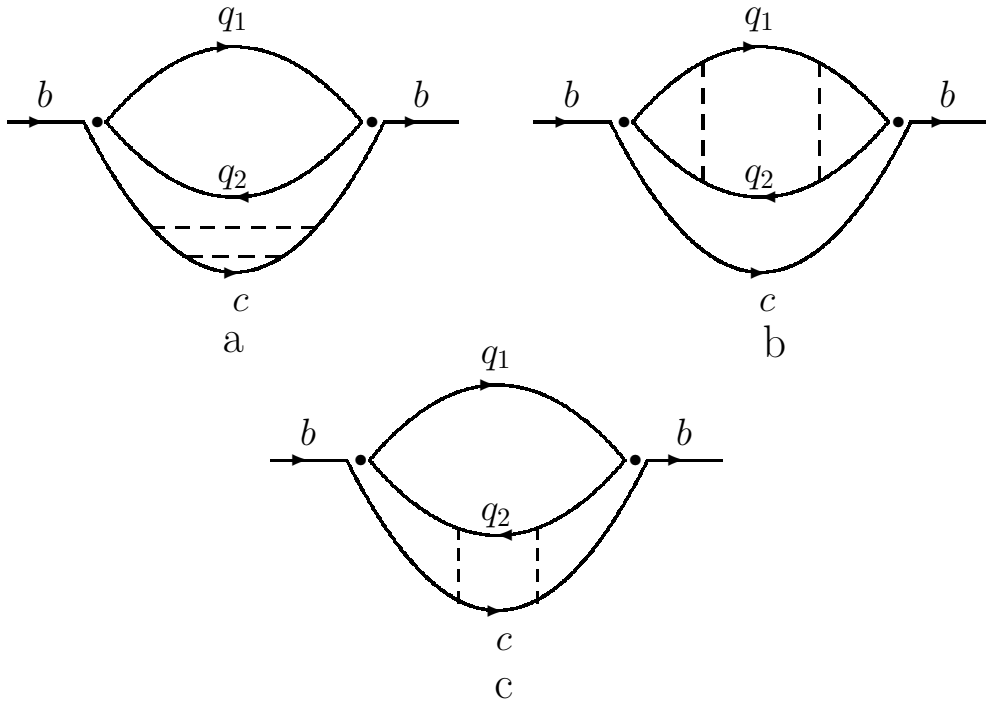

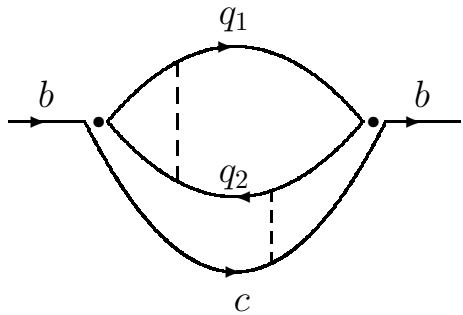
\begin{figure}
\unitlength 1.00mm
\begin{picture}(140.00,55.00)
\put(45.00,35.00){\vector(1,0){5.00}}
\put(50.00,35.00){\line(1,0){5.00}}
\put(95.00,35.00){\vector(1,0){5.00}}
\put(100.00,35.00){\line(1,0){5.00}}
\bezier{356}(55.00,35.00)(75.00,-5.00)(95.00,35.00)
\bezier{208}(58.00,35.00)(75.00,15.00)(92.00,35.00)
\bezier{208}(58.00,35.00)(75.00,55.00)(92.00,35.00)
\put(74.00,45.00){\vector(1,0){1.00}}
\put(76.00,25.00){\vector(-1,0){1.00}}
\put(74.00,15.00){\vector(1,0){2.00}}
\put(50.00,37.00){\makebox(0,0)[cb]{\large $b$}}
\put(100.00,37.00){\makebox(0,0)[cb]{\large $b$}}
\put(76.00,12.00){\makebox(0,0)[ct]{\large $c$}}
\put(75.00,48.00){\makebox(0,0)[cb]{\large $q_1$}}
\put(75.00,27.00){\makebox(0,0)[cb]{\large $q_2$}}
\put(57.00,35.00){\circle*{1.50}}
\put(93.50,35.00){\circle*{1.50}}
\multiput(80.00,16.00)(0,2.80){4}{\line(0,1){1.60}}
\multiput(67.50,27.00)(0,3.50){5}{\line(0,1){2.00}}
\end{picture}
\caption{One of the types of graphs of order $\as^2$, 
which do not contribute to the decay 
rate of $b \to c \overline q_2 q_1$,  because an overall color flows
into the colorless quark loop.} 
\end{figure}

\end{document}